\documentclass[12pt]{article}
\usepackage{amssymb}
\usepackage{epsfig}
\begin{document}
%
\setlength{\baselineskip}{0.65cm}
\setlength{\parskip}{0.35cm}
%
\begin{titlepage}
%
\begin{flushright}
ADP-01-61/T501\\
December 2001
\end{flushright}

\vspace*{2.0cm}
\begin{center}
\LARGE
\hbox to\textwidth{\hss
{\bf {Next-to-Leading Order QCD Corrections to the}} 
\hss}

\vspace*{0.5cm}
\hbox to\textwidth{\hss
{\bf {Polarized Hadroproduction of Heavy Flavors}}
\hss}

\vspace*{2.5cm}
\large 
{Ingo Bojak$^{a}$ and Marco Stratmann$^{b}$}

\vspace*{1.5cm}
\normalsize
{\em $^a$Special Research Centre for the Subatomic Structure of Matter,\\
Adelaide University, Adelaide SA 5005, Australia}\\

\vspace*{0.5cm}
{\em $^b$Institut f\"{u}r Theoretische Physik, Universit\"{a}t Regensburg,\\
D-93040 Regensburg, Germany}
\end{center}

\vspace*{2.5cm}
\begin{abstract}
\noindent
We present the complete next-to-leading order QCD corrections to the
polarized hadroproduction of heavy flavors which soon will be studied
experimentally in polarized $pp$ collisions at the BNL RHIC in order
to constrain the polarized gluon density $\Delta g$.
It is demonstrated that the dependence on unphysical renormalization
and factorization scales is strongly reduced beyond the leading order.
The sensitivity of the heavy quark spin asymmetry to $\Delta g$ 
is studied, including the limited detector acceptance for
experimentally observable leptons from heavy quark decays at the BNL RHIC.
As a further application of our results, gluino pair production 
in polarized $pp$ collisions is briefly discussed.
\end{abstract}
\end{titlepage}
\newpage
%
%
\section{Introduction and Motivation}
\noindent
Triggered by the measurement of the proton's spin-dependent deep-inelastic
structure function $g_1^p$ by the EMC~\cite{ref:emc}
more than a decade ago, combined experimental and theoretical efforts 
have led to an improved understanding of the spin structure of the
nucleon. In particular, we have gained some fairly precise information
concerning the total quark spin contribution to the nucleon spin.
The most prominent ``unknown'' is the elusive, yet unmeasured 
spin-dependent gluon density, $\Delta g$.
Hence current and future experiments designed to
further unravel the spin structure of the nucleon focus strongly on the issue
of constraining $\Delta g$.
In particular, information will soon be gathered for the first time 
at the BNL Relativistic Heavy-Ion Collider (RHIC)~\cite{ref:rhic}.

The main thrust of the RHIC spin program~\cite{ref:rhic} is to hunt down 
$\Delta g$ by measuring double spin asymmetries in longitudinally polarized $pp$
collisions at high energies. RHIC is particularly suited for this task, since the gluon
density is expected to participate dominantly in many different 
production processes. This is in contrast to deep-inelastic lepton-nucleon
scattering where the gluon enters only as a small correction 
in the next-to-leading order (NLO) of QCD and indirectly via the 
renormalization group evolution of the parton densities. Along with
the production of prompt photons and jets with high transverse momentum
$p_T$, heavy flavor pair creation is one of the most promising 
candidates at RHIC to study $\Delta g(x,\mu_F)$ over a broad range of the
momentum fraction $x$ and scale $\mu_F$. 
By virtue of the factorization theorem~\cite{ref:fact}, 
helicity densities extracted from different processes should match. 
This important universality property, exploited in ``global'' fits of
parton densities, can also be tested at RHIC.

In the lowest order (LO) in the strong coupling $\alpha_s$, heavy flavor 
pair production in hadron-hadron collisions proceeds through two 
parton-parton subprocesses,
\begin{equation}
\label{eq:loproc}
q\overline{q} \to Q\overline{Q}\;\;\;\;\mathrm{and}\;\;\;\;
g g\to Q\overline{Q}\;\;.
\end{equation}
Gluon-gluon fusion is known to be the by far dominant mechanism
for charm and bottom production in the unpolarized case in all
experimentally relevant regions of phase space 
\cite{ref:smith1,ref:nason,ref:nasontotal}.
This feature, true also in the polarized case
unless $\Delta g$ is exceedingly small, makes heavy quark 
production a particularly suited tool to study the gluon density.
However, NLO QCD corrections to the LO subprocesses in Eq.~(\ref{eq:loproc})
have to be included for a reliable description.
First and foremost this is due to the strong dependence of the LO
results on unphysical theoretical conventions such as the 
factorization scale, which reflects the amount of arbitrariness in the
separation of short- and long-distance physics.
In addition, the size of the NLO corrections
turns out be quite sizable and not uniform in, e.g., the $p_T$ of
one of the heavy quarks \cite{ref:smith1,ref:smith2}. 
The latter feature rules out the use of any approximations
as estimates of the complete NLO corrections. 
The computation of the NLO corrections are fairly involved since one
has to keep track of the mass of the heavy quark, $m$, throughout the
calculation. Massless approximations are bound to fail at small-to-medium
values of $p_T$ where $m\simeq{\cal{O}}(p_T)$ and the cross section is
large. So far only unpolarized NLO results had been available,
see Refs.~\cite{ref:smith1,ref:smith2,ref:nason} and
\cite{ref:nasontotal} for the differential and total cross sections, respectively.
Polarized LO expressions can be found in Refs.~\cite{ref:contolo,ref:karliner},
but the complete NLO results are presented for the first time in this work. 

Apart from calculational difficulties, a further complication arises when 
one tries to match theoretical parton-level results for heavy flavor
production rates with experimental ones. Experiments can only observe 
the remains of heavy quark (meson) decays -- usually leptons.
In practice they also have to impose
cuts on these particles to insure a proper $c$ and $b$ quark separation and
to take care of the, usually limited and non uniform, detector acceptance.
One thus has to find
a practical way to incorporate hadronization, lepton-level cuts,
and the detector acceptance in an analysis based on parton-level calculations, since
they can distort spin asymmetries if polarized and unpolarized cross sections
are affected differently. This may lead to incorrect conclusions
about $\Delta g$. Heavy flavor
decays usually have multi-body kinematics making it 
difficult if not impossible to trace back cuts to the parton-level analytically. 
Instead we propose to use ``efficiencies'', to be defined below, 
for bins in $p_T$ and pseudo-rapidity $\eta$ of the heavy quark, 
to model its hadronization, decay, and crucial detector features. 

Heavy flavor production at RHIC is also interesting 
for reasons other than $\Delta g$. There is a well-known, longstanding
discrepancy between data from the Tevatron collider for unpolarized open $b$ production
and theory \cite{ref:b-report}.
Recently, $b$ rates in $ep$ and $\gamma\gamma$ collisions were also 
found to be in excess of theoretical predictions \cite{ref:b-report}.
The fact that open $c$ production seems to be
fairly well described by theory makes these experimental findings 
even more puzzling, since perturbative QCD should be 
more reliable for the much heavier $b$ quark.
This discrepancy for inclusive $b$ production has revived 
speculations about new physics.
For instance, the parameters of the minimal supersymmetric extension of the 
standard model can be tuned such that relatively light gluinos 
$\tilde{g}$ exclusively decay into even lighter
(s)bottom quarks $\tilde{b}$ and $b$'s: $\tilde{g}\to b\tilde{b}$
\cite{ref:lightgluino}. In this way the yield of $b$'s
is enhanced, and the Tevatron data can be reproduced while still complying 
with direct supersymmetry searches and precision measurements in $e^+e^-$ 
collisions \cite{ref:lightgluino}.
As will be briefly discussed below, our LO and NLO results for the dominant
gluon-gluon fusion subprocess also contain the production of 
gluino pairs, $gg\rightarrow \tilde{g}\tilde{g}$,
after adjusting the color factors appropriately\footnote{The
$q\overline{q}\rightarrow \tilde{g}\tilde{g}$ 
subprocess receives new contributions absent in 
$q\overline{q}\rightarrow Q\overline{Q}$.} \cite{ref:smith1}. 
Thus one can also estimate the spin-dependent hadroproduction 
of (light) gluinos at RHIC and at a conceivable polarized version of the 
CERN LHC in the distant future.
Other explanations \cite{ref:ktfact} of the observed 
$b$ excess make use of unintegrated gluon distributions and 
$k_T$-factorization, however, NLO calculations in this
framework are still lacking.
If results from the just upgraded Tevatron
confirm the $b$ excess, RHIC will play an important r{\^o}le 
in deciphering the underlying mechanism since it
allows to study the energy- {\em and} spin-dependence of both $c$ and $b$
production.

Finally, our results are also required for a fully consistent
description of the polarized photoproduction of heavy quark pairs. 
Apart from the ``direct'' process $\gamma g\to Q\overline{Q}$,
where the NLO corrections have been calculated 
in Refs.~\cite{ref:photo,ref:contophot}, the (quasi-)real photon 
can also resolve into its hadronic content before the hard scattering takes place.
The introduction of photonic parton densities is mandatory for
the factorization of mass singularities of the direct process
associated with collinear $\gamma\to q\overline{q}$, $q=u,d,s$, splitting.
Polarized ``resolved'' photon processes, Eq.~(\ref{eq:loproc}), 
have been estimated \cite{ref:resolved} 
to be small for fixed target experiments like COMPASS \cite{ref:compass}, but can be
significant at proposed future polarized lepton-hadron
colliders like the EIC \cite{ref:eic}.

In this paper we focus on the outline and main results of our calculation.
In Section~2 we give a brief description of the calculational steps and methods
we have employed, in particular of the subtleties which arise due to the
spin-dependence. In Section~3 we first discuss the different features of
the NLO corrections to the polarized gluon-gluon fusion subprocess, always 
in comparison to the unpolarized case.
Next we demonstrate the significantly reduced dependence on unphysical 
renormalization and factorization scales in NLO QCD for heavy flavor production at RHIC. 
We finish with studying the sensitivity of the heavy quark spin asymmetry
to $\Delta g$ at RHIC energies including realistic cuts on experimentally observable
leptons from heavy quark decays. We conclude in Section~4.
Technical details and lengthy analytical results are omitted throughout and
will be presented elsewhere \cite{ref:upcoming} along with 
further phenomenological applications. 

%
\section{Outline of the Technical Framework}
\noindent
The ${\cal{O}}(\alpha_s^3)$ NLO QCD corrections 
to heavy flavor production comprise of three parts:
the one-loop virtual corrections to the LO processes in Eq.~(\ref{eq:loproc}),
the real ``$2\to 3$'' 
corrections to Eq.~(\ref{eq:loproc}) with an additional gluon in the
final state, and a new production mechanism, $gq (\overline{q}) \to Q\overline{Q} q (\overline{q})$,
appearing for the first time at the NLO level.
We choose the well-established framework of $n$-dimensional regularization, 
with $n=4+\varepsilon$, to tame the singularities of the loop- 
and $2\to 3$ phase space integrals.
Ultraviolet singularities show up only in the virtual corrections and are removed
by on-shell mass and coupling constant renormalization
at a scale $\mu_R$. The latter is performed in
a variant of the $\overline{\mathrm{MS}}$ scheme which is usually adopted 
for heavy flavor production \cite{ref:nason,ref:nasontotal,ref:smith2}.
This prescription is characterized by the decoupling
of heavy quark loop contributions to the gluon self energy and
leads to a fixed flavor scheme with $n_{lf}=n_f-1$ light flavors
active in the running of $\alpha_s$ and in the scale $\mu_F$ evolution of the parton
densities. Infrared (IR) divergencies of the virtual diagrams are cancelled by the
soft poles of the $2\to 3$ contributions. This includes also
double, $1/\varepsilon^2$, pole terms which show up when IR and
mass/collinear (M) singularities coincide. 
The left over $1/\varepsilon$ M singularities are then absorbed into the
bare parton densities by the standard factorization procedure 
in the $\overline{\mathrm{MS}}$ scheme.
Other renormalization/factorization schemes can be 
obtained easily by additional finite scheme transformations.

The required squared matrix elements $|M|^2$ for both unpolarized and
longitudinally polarized processes are obtained {\em simultaneously} by calculating
them for arbitrary helicities $\lambda_{1,2}=$``$\pm$'' 
of the incoming quarks or gluons, i.e.,
\begin{equation}
\label{eq:punp}
|M|^2(\lambda_1,\lambda_2)=\overline{|M|}^2+\lambda_1\lambda_2\,
\Delta |M|^2\,,
\end{equation}
using the standard helicity projection operators for bosons and fermions
(see, e.g., Ref.~\cite{ref:craigie}). Results obtained 
for the unpolarized $\overline{|M|}^2$ can be compared to the 
literature \cite{ref:smith1,ref:nason,ref:nasontotal,ref:smith2} 
which serves as an important consistency check for the correctness of
our new helicity dependent results $\Delta |M|^2$. To facilitate this comparison 
we closely follow the calculational steps and methods adopted in
\cite{ref:smith1,ref:smith2}. It should be noted that,
contrary to the unpolarized case \cite{ref:smith2},
the processes $q\overline{q} \to Q\overline{Q}g$ and $gq \to Q\overline{Q} q$ 
are not related by crossing for polarized initial states and have to be
calculated separately. We should also recall here the definition of 
the spin-dependent parton densities,
\begin{equation}
\label{eq:pdfdef}
\Delta f(x,\mu_F) \equiv f_+^+(x,\mu_F) - f_-^+(x,\mu_F)
\end{equation}
where $f_+^+$ $(f_-^+)$ denotes the probability to find a parton 
$f=q,\,\overline{q},\,g$ at a scale $\mu_F$ with momentum fraction $x$
and helicity $+$ ($-$) in a proton with helicity $+$ [the unpolarized
parton densities $f(x,\mu_F)$ are obtained by taking the sum in
(\ref{eq:pdfdef})]. In the following the compact notation
$\tilde{\phi}$ denotes both an unpolarized
quantity $\phi$ and its longitudinally polarized 
analogue $\Delta \phi$.

The virtual (V) cross section for the $q\overline{q}$ and $gg$ initial states 
is obtained up to ${\cal{O}}(\alpha_s^3)$ only from the interference 
between the virtual and Born amplitudes. 
Loop momenta in the numerator are dealt with by applying an adapted version
of the Passarino-Veltman reduction program to scalar integrals \cite{ref:pvdecomp},
which properly accounts for all possible $n$-dimensionally regulated 
divergencies in QCD. The required scalar integrals can be found in
\cite{ref:smith1}, however, we have checked them by
standard Feynman parameterization techniques (see also \cite{ref:ingophd}).
The renormalized, color-averaged results can be decomposed 
according to their color structure as
\begin{eqnarray}
\label{eq:mvqq}
|\tilde{M}_{q\overline{q}}|^2_{\mathrm{V}} \!\!\!&=&\!\!\! 
g^6 \frac{C_F}{4N_C} \left[ 2 C_F \tilde{N}_{QED} + C_A \tilde{N}_{OK} +
\tilde{N}_{QL} \right],
\\
\label{eq:mvgg}
|\tilde{M}_{gg}|^2_{\mathrm{V}} \!\!\! &=& \!\!\!g^6
\tilde{E}_{\varepsilon}^2 \frac{1}{2(N_C^2-1)}
\left[ (2C_F)^2 \tilde{U}_{QED} + C_A^2 \tilde{U}_{OQ} +
\tilde{U}_{KQ} + 2 C_F \tilde{U}_{RF} + C_A \tilde{U}_{QL} \right],
\end{eqnarray}
where $g^2=4\pi \alpha_s$, $E_{\varepsilon}=1/(1+\varepsilon/2)$, 
and $\Delta E_{\varepsilon}=1$. The color factors are expressed
in terms of the Casimir operators $C_F=(N_C^2-1)/2N_C$ and $C_A=N_C$,
where $N_C$ denotes the number of colors.
The lengthy expressions for $\tilde{N}_i$ and $\tilde{U}_i$ 
in Eqs.~(\ref{eq:mvqq}) and (\ref{eq:mvgg}) can be found
in \cite{ref:upcoming}. For the $gg$ process we have a slightly different
way of splitting up the results according to color than Ref.~\cite{ref:smith1}.
The choice in Eq.~(\ref{eq:mvgg}) ensures that the ``Abelian'' $\tilde{U}_{QED}$
is identical to the QED part of $\gamma g\to Q\overline{Q}$ in
Ref.~\cite{ref:photo} after taking into account the usual factor $1/(2N_C)$ 
for replacing a photon by a gluon. Furthermore, 
compared to Ref.~\cite{ref:smith1} an additional function
$\tilde{U}_{RF}$ appears in Eq.~(\ref{eq:mvgg}) since we are interested in
the general case $\mu_R\neq\mu_F$.
Our unpolarized results fully agree with the 
corresponding expressions in \cite{ref:smith1,ref:smith2} except for
$U_{QL}$ in Eq.~(6.22) of Ref.~\cite{ref:smith1} which contains a
numerically irrelevant misprint\footnote{We thank J.\ Smith for his help in
clarifying this issue.}.

The real $2\to 3$ gluon bremsstrahlung corrections (R)
can be split up in a similar way according to color structure and
read
\begin{equation}
\label{eq:mrgg}
|\tilde{M}_{gg}|^2_{\mathrm{R}} = g^6
\tilde{E}_{\varepsilon}^2 \frac{1}{2(N_C^2-1)}
\left[ (2C_F)^2 \tilde{D}_{QED} + C_A^2 \tilde{D}_{OQ} +
\tilde{D}_{KQ}\right]\;\;,
\end{equation}
and accordingly for the $q\overline{q}$ and $gq$ processes \cite{ref:upcoming}.
In order to isolate the divergencies of (\ref{eq:mrgg}) appearing in the
soft limit, which cancel the remaining singularities of the virtual
matrix elements, we ``slice'' the  $2\to 3$ result into a
``soft'' (S) and a ``hard'' (H) gluon part by introducing a small auxiliary
quantity $\Delta$ \cite{ref:smith1,ref:smith2}. The kinematics of the
soft gluon cross section is effectively that of a $2\to 2$
process, and all phase space integrations can be performed easily.
Upon combination with the virtual cross section the $1/\varepsilon$
IR and $1/\varepsilon^2$ IR+M
singularities, all proportional to the $n$-dimensional Born
cross section, drop out. Thus the ``soft+virtual'' (S+V) cross section
becomes finite, except for $1/\varepsilon$ M singularities that cancel
against ``soft'' $x=1$ contributions, see Eq.~(\ref{eq:scheme}),
in the mass factorization procedure discussed below.
Phase space integrations for the hard part are more
subtle and require some care. As in \cite{ref:smith1,ref:smith2} we
are interested in the double differential 
single inclusive cross section for the production of a heavy quark (or antiquark). 
For stable numerical simulations later on it is advantageous
to perform the integrations over the phase space of the two not
observed partons analytically as far as possible. To achieve this
requires extensive partial fractioning to reduce all phase space
integrals to a standard form \cite{ref:smith1,ref:photo,ref:ingophd}.
A sufficient set of four- and $n$-dimensional integrals are again
conveniently collected in \cite{ref:smith1}, but we have recalculated 
and confirmed this set. Analytical results for the S+V cross sections will be given in
\cite{ref:upcoming}; the final expressions for the hard part are
too lengthy but can be found in our computer code.

To obtain a finite result, all remaining M singularities
have to be removed by the standard mass factorization procedure
which makes use of the fact that any collinear singular partonic
cross section $d\tilde{\sigma}_{ij}^M$ can be schematically 
written as
\begin{equation}
\label{eq:coll}
d\tilde{\sigma}_{ij}^M(k_1,k_2)=\int_0^1 dx_1 \int_0^1 dx_2\,
\tilde{\Gamma}_{li}(x_1,\mu_F/\mu)\, \tilde{\Gamma}_{mj} (x_2,\mu_F/\mu)
\, d\tilde{\hat{\sigma}}_{lm}(x_1k_1,x_2 k_2)\;\;,
\end{equation}
where $k_{1,2}$ are the momenta of the incoming partons.
Here $d\tilde{\hat{\sigma}}_{lm}$ denotes the {\em finite}
partonic cross section, $\mu$ is the scale introduced to render
the coupling $g$ dimensionless in $n$ dimensions, and
the transition functions $\Gamma_{ij}$ are given up to NLO by
\begin{equation}
\label{eq:scheme}
\tilde{\Gamma}_{ij}(x,\mu_F/\mu) = \delta_{ij}\,\delta(1-x)+
\frac{\alpha_s}{2\pi}
\left[ \tilde{P}_{ij}^{(0)}(x) \left( 
\frac{2}{\varepsilon}+\gamma_E-\ln 4\pi + \ln \frac{\mu_F^2}{\mu^2}\right) +
\tilde{\xi}_{ij}(x) \right]
\end{equation}
with $\tilde{P}_{ij}^{(0)}$ the LO splitting functions. Therefore all
M singularities, Eq.~(\ref{eq:coll}), can be absorbed {\em universally}
into the bare parton densities upon calculating physical, i.e., hadron-level, observables.
The choice of the factorization scale $\mu_F$ and the arbitrary functions
$\tilde{\xi}_{ij}$ reflects the amount of arbitrariness in the separation of 
short-distance and long-distance physics. In the $\overline{\mathrm{MS}}$ scheme,
which we choose, one has $\tilde{\xi}_{ij}=0$.

Two subtleties, which show up in $n$-dimensionally regulated
spin-dependent calculations beyond the LO of QCD, 
have to be mentioned here. First and foremost, the projection operators onto certain
helicity states, i.e., $\epsilon_{\mu\nu\rho\sigma}$ and
$\gamma_5$, are of purely four dimensional nature, and there exists no
straightforward and unique generalization to $n\neq 4$ dimensions.
We treat them by applying the internally consistent HVBM prescription 
\cite{ref:hvbm}, where the Levi-Civita $\epsilon$-tensor continues to be a 
genuinely four-dimensional object. $\gamma_5$ is defined as in four dimensions, implying 
$\{\gamma_\mu,\gamma_5\}=0$ for $\mu=0,1,2,3$ 
and $[\gamma_\mu,\gamma_5 ]=0$ otherwise. 
The price to pay are $(n-4)$ dimensional scalar products, usually
denoted by $\widehat{k\cdot p}$ (``hat momenta'') appearing
alongside the usual $n$-dimensional scalar products.
In our case only a single hat momenta combination $\hat{p}^2=-\widehat{p\cdot p}$ 
appears in the polarized $2\to 3$ cross section
and can be accounted for by an appropriately modified phase space
formula \cite{ref:photo}. These contributions are inherently of
${\cal{O}}(\varepsilon)$ and only contribute to the final result
when they pick up a $1/\varepsilon$ pole.
Secondly, the unphysical helicity violation at the $qqg$ vertex in the HVBM
scheme, which is reflected by $P_{qq}^{(0)} \neq \Delta P_{qq}^{(0)}$ in
$n$ dimensions, has to be undone by an additional finite renormalization
\cite{ref:werner} $\Delta \xi_{qq}=-4C_F (1-x)$ in Eq.~(\ref{eq:scheme})
in the conventional $\overline{\mathrm{MS}}$ scheme. Only then 
$d\Delta\hat{\sigma}_{q\overline{q}}=-d\hat{\sigma}_{q\overline{q}}$ is obtained, the
result which is expected due to helicity conservation.

%
\section{Numerical Results and Phenomenological Aspects}
\noindent
Before presenting results for hadronic heavy flavor distributions we first
discuss the {\em total partonic} subprocess cross sections 
$\tilde{\hat{\sigma}}_{ij}$, $i, j=q, \overline{q}, g$. 
They can be expressed in terms of LO and NLO functions $\tilde{f}^{(0)}_{ij}$ and 
$\tilde{f}^{(1)}_{ij}$, $\;\tilde{\!\!\bar{f}}{}_{ij}^{(1)}$, respectively, 
which depend only on a single {\em scaling} variable $\eta=s/(4m^2)-1$:
%
\begin{figure}[th]
\epsfxsize=\textwidth 
\epsfbox{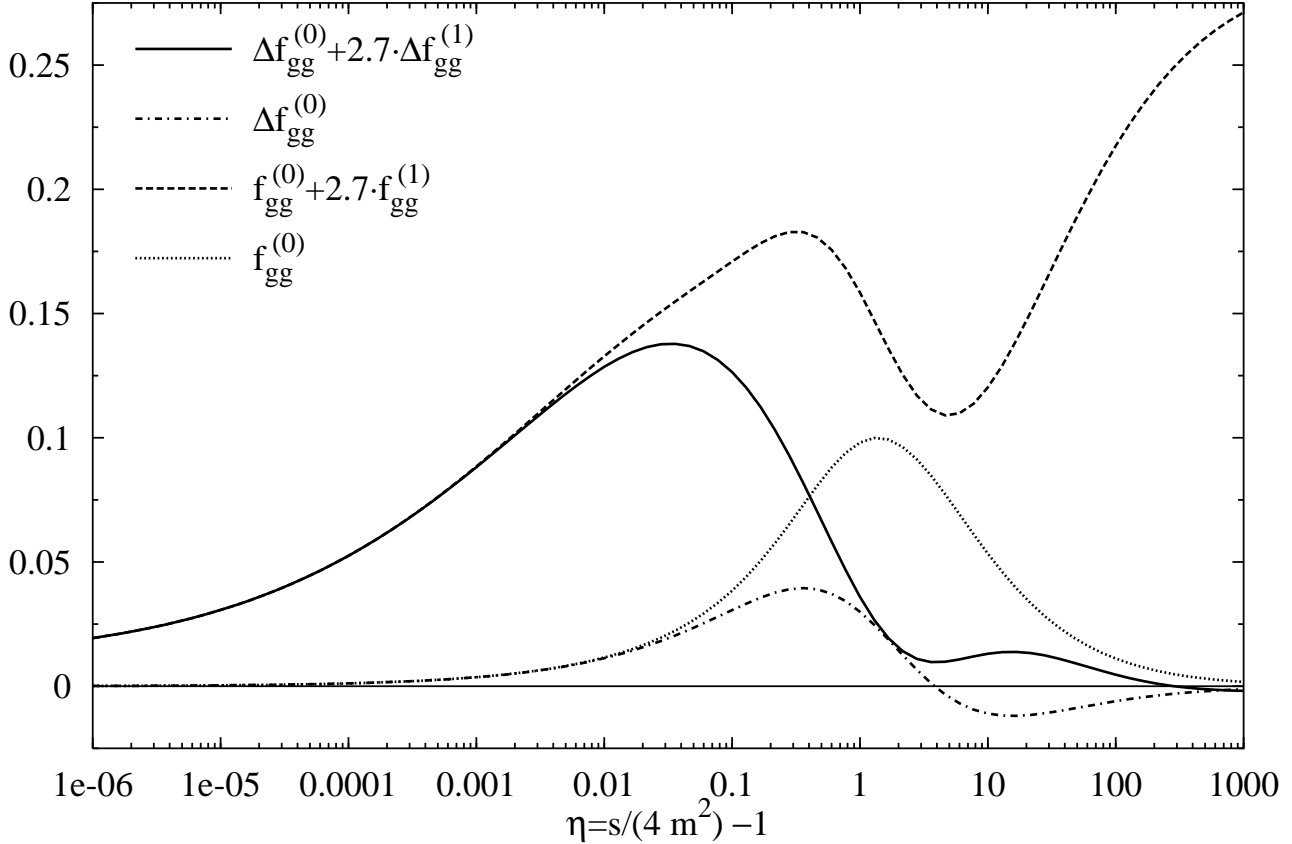}
\caption{\label{fig:gg}\sf
$(m^2/\alpha_s^2) \tilde{\hat{\sigma}}_{gg}$ in NLO ($\overline{\mathrm{MS}}$)
and LO as a function of $\eta$
according to Eq.~(\ref{eq:totalpartonic}), where 
we have set $\mu_F=\mu_R=m$ for simplicity and
$4\pi\alpha_s=2.7$ as appropriate for charm production.}
\end{figure}
\begin{equation}
\label{eq:totalpartonic}
\tilde{\hat{\sigma}}_{ij}(s,m^2,\mu_F,\mu_R) =\frac{\alpha_s^2}{m^2} \left\{
\tilde{f}^{(0)}_{ij}(\eta) + 4\pi \alpha_s \left[
\tilde{f}^{(1)}_{ij}(\eta) + \;\tilde{\!\!\bar{f}}{}_{ij}^{(1)}(\eta) \ln \frac{\mu_F^2}{m^2}
+ \frac{\beta_0}{8\pi^2} \tilde{f}^{(0)}_{ij}(\eta) \ln \frac{\mu_R^2}{\mu_F^2}
\right]\right\}\;,
\end{equation}
where $\alpha_s=\alpha_s(\mu_R^2)$ and $\beta_0=(11C_A-2 n_{lf})/3$.
Hence the $\tilde{\hat{\sigma}}_{ij}$ are particularly suited for studying  
the main features of the NLO corrections in the most transparent way.
The $\tilde{f}_{ij}^{(1)}$ are non-trivial functions of $\eta$ and can 
be easily obtained from our double differential analytical results for the 
partonic cross sections by numerical integrations. In the unpolarized case 
they have been cast into a compact semi-analytical form \cite{ref:nasontotal} for fast 
numerical calculations of the total hadronic heavy flavor cross section 
which is a desirable future project in the polarized case.
The $\;\tilde{\!\!\bar{f}}{}_{ij}^{(1)}$ can be derived just from mass factorization,
since terms proportional to $\ln \mu_F^2$ originate only from Eq.~(\ref{eq:scheme}).
The last term in Eq.~(\ref{eq:totalpartonic}) vanishes for the standard choice
$\mu_F=\mu_R$. In NLO this term follows straightforwardly from the LO result by replacing
$\alpha_s \to \alpha_s \left(1+\alpha_s \frac{\beta_0}{4\pi} 
\ln \frac{\mu_R^2}{\mu_F^2}\right)$ thanks to the renormalization group invariance
of the cross section.

In Fig.~\ref{fig:gg} we present the gluon-gluon subprocess cross section 
$(m^2/\alpha_s^2) \tilde{\hat{\sigma}}_{gg}$ in LO and NLO
for $\mu_F=\mu_R=m$ as a function of $\eta$ in the $\overline{\mathrm{MS}}$ scheme.
The threshold for $Q\overline{Q}$ production, $s=4m^2$, is located at $\eta=0$.
It turns out that the NLO corrections are significant in the entire $\eta$ range.
At threshold the polarized and unpolarized cross sections are equal, thus
Eq.~(\ref{eq:punp}) implies that $|M_{gg}|^2(+-)\to 0$ 
as $\eta \to 0$. Unlike in LO where $\tilde{\hat{\sigma}}_{gg}$ approaches zero at threshold, 
it tends to a constant in NLO, 
$\frac{\alpha_s^3}{8m^2}\frac{1}{2(N_C^2-1)} \left[(2C_F)^2-C_A^2+\frac{5}{2}\right]\pi^2$,
due to the ``Coulomb singularity'' present in the S+V part.
It should be noted that in the threshold region logarithms 
from soft gluon emissions also contribute significantly even
at the lowest $\eta$ shown. 
In the high energy domain, $\eta\to\infty$,
our polarized and unpolarized results behave rather differently. 
Here Feynman diagrams with a gluon exchange in the $t$-channel 
drive the unpolarized NLO result to a plateau value \cite{ref:smith1} whereas the
polarized NLO cross section vanishes like the LO one, i.e., 
$|M_{gg}|^2(++)\to |M_{gg}|^2(+-)$ in Eq.~(\ref{eq:punp}) 
as $\eta\to\infty$. Similar observations have been made in the photoproduction 
case $\gamma g\to Q\overline{Q}$ \cite{ref:photo}. 
The scaling function $\tilde{f}_{q\overline{q}}$ fulfils $\Delta f_{q\overline{q}}=-f_{q\overline{q}}$
after taking into account the additional finite factorization mentioned above
to restore helicity conservation at the $qqg$ vertex. 
The behaviour of $f_{q\overline{q}}$ was discussed in \cite{ref:smith2} and shall not
be repeated here.
The genuine NLO scaling function $\tilde{f}_{gq}$ is numerically much smaller than
$\tilde{f}_{gg}$ and can be found in \cite{ref:upcoming} but exhibits the same 
high-energy $\eta\to\infty$ behaviour as $\tilde{f}_{gg}$, 
i.e., $f_{gq}$ approaches a plateau while $\Delta f_{gq}$ vanishes.

The physical, i.e., experimentally observable, total cross section is obtained 
by convoluting the partonic cross sections in Eq.~(\ref{eq:totalpartonic}) with the
appropriate flux of parton densities evolved to the scale $\mu_F$,
\begin{equation}
\label{eq:totalxsec}
\tilde{\sigma}(S,m^2,\mu_F,\mu_R)=
\sum_{ij} \int_{\frac{4m^2}{S}}^1 dx_1 \int_{\frac{4m^2}{x_1S}}^1 dx_2\,
\tilde{f}_i(x_1,\mu_F) \, \tilde{f}_j(x_2,\mu_F) \, 
\tilde{\hat{\sigma}}_{ij}(s,m^2,\mu_F,\mu_R)\;\;,
\end{equation}
where $S$ is the available hadron-hadron c.m.s.\ energy and $s=x_1x_2 S$.
In a similar fashion differential heavy (anti-)quark inclusive distributions
like $d^2\tilde{\sigma}/dp_Tdp_L$ can be derived by convolution with our double 
differential partonic cross sections.
It also should be kept in mind that beyond the LO of QCD, parton densities and partonic
cross sections have to be taken in the same factorization scheme
in order to guarantee that Eq.~(\ref{eq:totalxsec}) is independent of
unphysical theoretical conventions up to the order in $\alpha_s$ considered in the
calculation. 

\begin{figure}[t]
\epsfxsize=\textwidth 
\epsfbox{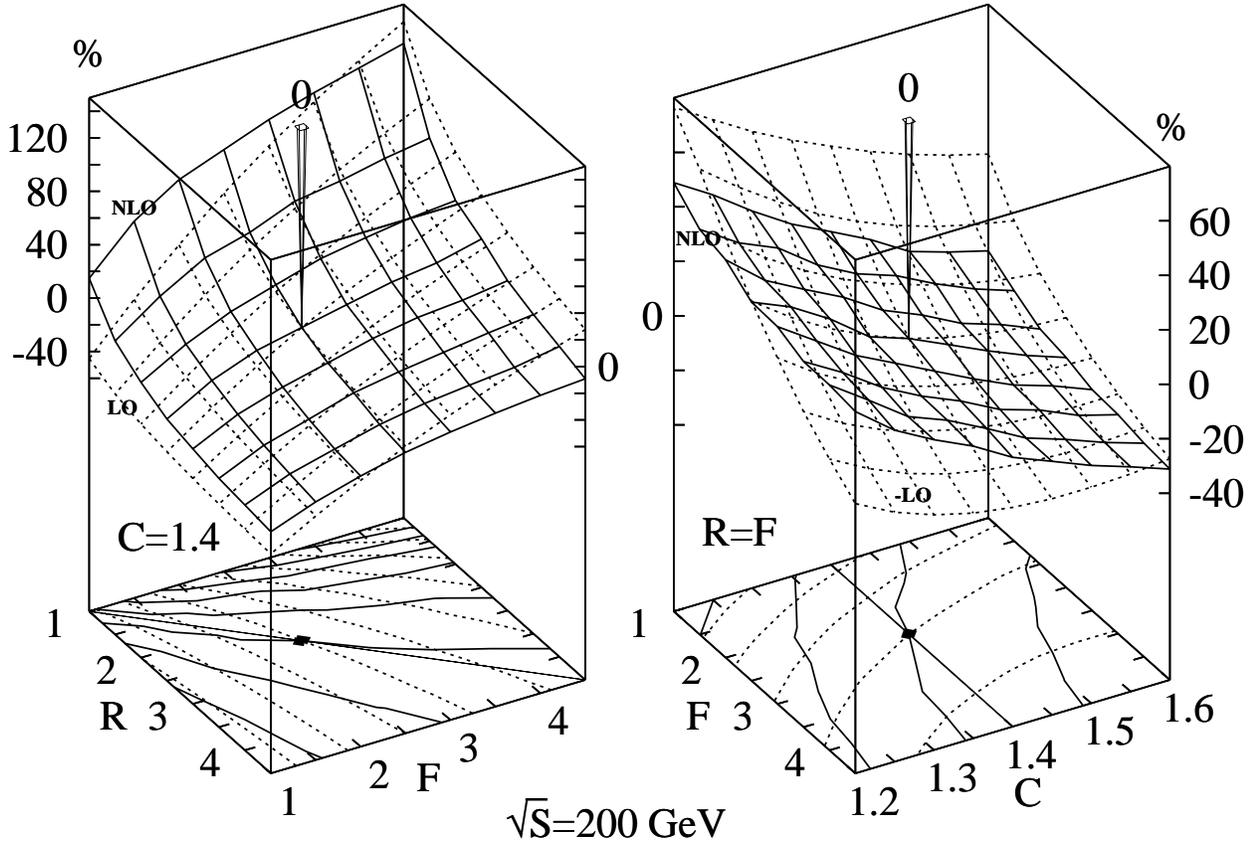}
\caption{\label{fig:rfc}\sf
Deviation [in $\%$] of the polarized total charm cross section 
in LO (dotted) and NLO (solid) from a reference choice (``0-pin'' marker, see text)
-- {\bf left part}: as a function of $\mu_{F}$ and $\mu_{R}$
for fixed $m$; {\bf right part}: as a function of
$\mu_F$ and $m$ with $\mu_R=\mu_F$, here the LO result is multiplied by a
factor (-1).
Corresponding contour lines in steps of $20\%$ are given at the base of each plot.}
\end{figure}
One of the main motivations of performing the NLO calculations
was to reduce the dependence on the choice of $\mu_F$ and $\mu_R$ which is 
completely arbitrary in LO and can lead to sizable ambiguities in predictions
for $\tilde{\sigma}(S,m^2)$ and the corresponding spin asymmetry to be defined
below. In Fig.~\ref{fig:rfc} we demonstrate that the NLO results for the polarized 
charm production cross section are indeed more robust under scale variations 
than LO estimates. In the left panel of Fig.~\ref{fig:rfc} we vary $\mu_F$ and $\mu_R$ 
independently of each other in the range $\mu_R^2=R\, m^2$ and $\mu_F^2=F\, m^2$ 
with $1\leq R,F\leq 4.5$ for fixed $m=1.4\,\mathrm{GeV}$
at a typical RHIC energy of $\sqrt{S}=200\,\mathrm{GeV}$
using the GRSV ``standard'' set of polarized parton densities \cite{ref:grsv}.
In the right part of Fig.~\ref{fig:rfc} we employ the conventional choice $\mu_R\equiv\mu_F$
and vary $\mu_F$ and $m=C\,\mathrm{GeV}$ in a typical range for the charm pole mass. 
In order to better visualize the uncertainties due to scale and mass variations we show
$\Delta\sigma (R,F,C)/\Delta\sigma (R=2.5,F=2.5,C=1.4)-1$, i.e., the deviation
in percent of the total polarized charm production cross section according to 
Eq.~(\ref{eq:totalxsec}) for variable $\mu_{F,R}$ and $m$
with respect to a reference cross section taken at fixed $\mu_F^2=\mu_R^2=2.5m^2$
with $m=1.4\,\mathrm{GeV}$. To better guide the eye, 
contour lines in steps of $20\%$ are plotted at the base of each plot.
Here we also indicate the common choice $\mu_R=\mu_F$ and
$m=1.4\,\mathrm{GeV}$ (thin solid lines) in the left and right part, respectively.

The NLO result in the left part of Fig.~\ref{fig:rfc} is considerably ``flatter''
than the LO result with respect to variations of $\mu_F$ but shows, however, 
slightly more variation with $\mu_R$. Not unexpectedly and more importantly 
it turns out that the usual choice $\mu_R=\mu_F$ almost coincides with the contour line
of zero deviation from the reference cross section in NLO, in stark contrast to
the situation at LO. This leads to the improved stability of the NLO prediction
in the right panel of Fig.~\ref{fig:rfc} for variations of a {\em common} scale
$\mu_F\equiv\mu_R$ at a given charm mass $m$.
Here variations of the charm mass cause the major uncertainty of about $\pm 30\%$
in the NLO predictions. In LO we find considerable uncertainty stemming from 
variations of $\mu_F$ on top of that. 
It should be also noted that qualitatively similar results are
obtained for $\sqrt{S}=500\,\mathrm{GeV}$ and bottom production at RHIC.
Usually in NLO the terms proportional
to $\ln \mu_F^2/m^2$ and $\ln \mu_R^2/\mu_F^2$ in Eq.~(\ref{eq:totalpartonic})
start to have a compensating effect for different choices for $\mu_F$ and
$\mu_R$ and also provide some guidance 
that $\mu_f\sim{\cal{O}}(m)$ and $\mu_F\sim\mu_R$ in order
to avoid large logarithms in the hard partonic cross sections.
Ultimately one expects the dependence on $\mu_F$ and $\mu_R$ to be reduced more and more 
if higher and higher orders in $\alpha_s$ are considered. However, as was briefly
explained above in connection with Fig.~\ref{fig:gg}, in the reaction studied here 
new types of Feynman diagram topologies enter the calculation for the first time at the NLO
level, whereas in next-to-NLO (NNLO) and beyond no qualitatively different
diagrams appear. Hence in a sense NLO is the first ``complete'', 
non-trivial order of perturbation theory for heavy flavor production, and it is 
pleasing that scale stability improvements nevertheless clearly set in 
without considering NNLO corrections which seem unattainable at this time.

Instead of measuring polarized cross sections like
$\Delta \sigma(S,m^2)$ directly, experiments will usually
study the related longitudinal spin asymmetry defined by
\begin{equation}
\label{eq:asym}
A(S,m^2) \equiv \frac{\Delta\sigma(S,m^2)}{\sigma(S,m^2)}
\end{equation}
in case of the total cross section and accordingly for differential
heavy quark distributions.
The experimental advantage of this quantity is that one does not need
to determine the absolute normalization of the cross section 
$\Delta\tilde{\sigma}(S,m^2)$ which is usually difficult to obtain. However, one should
keep in mind that the situation in the unpolarized case is far from clear, in particular
concerning bottom, and hence it would be reasonable to determine the unpolarized
and polarized cross section separately. We note that for small variations of the
scales the relative deviation of the asymmetry can be written as
$\frac{\delta A}{A}=\frac{\delta\Delta\sigma}{\Delta\sigma}-\frac{\delta\sigma}{\sigma}$.
It turns out for the variations of $\mu_F$, $\mu_R$, and $m$
considered above that $\frac{\delta\Delta\sigma}{\Delta\sigma}$ and
$\frac{\delta\sigma}{\sigma}$ are almost equal in NLO, whereas they can differ strongly
in LO. As a result it is even more true for the asymmetry that NLO 
results are highly stable,
whereas the LO uncertainty is huge, in particular for the choice $\mu_R\equiv\mu_F$.
We will explore this in detail in \cite{ref:upcoming}, but wish to point out here
that LO determinations of $\Delta g$ using the asymmetry alone will necessarily have
a prohibitively large theoretical error, a NLO analysis is a must in that case.

Finally, let us turn to the important question of whether heavy flavor production
at RHIC can be used to discriminate between different polarized gluon densities.
To address this question thoroughly one has to take into account an estimate of
the statistical significance of a measurement of a heavy quark spin asymmetry at RHIC. 
Compared to direct photons or jets which are directly observed in the detector this
is a rather involved problem for heavy flavors. With the PHENIX detector at RHIC
charm and bottom quarks can be identified only through their decay products,
preferably leptons. However, the electron and muon detection is rather limited in 
pseudo-rapidity,
$|\eta_{e}|\leq 0.35$ and $1.2\leq |\eta_{\mu}|\leq 2.4$, respectively,
and cuts in the lepton $p_T$ have to be imposed in order to separate charm and bottom.
Since heavy flavor decays usually have a multi-body kinematics and may proceed through cascades,
cuts on the observed leptons are difficult to translate back to the 
calculated parton, i.e., heavy quark, level. One possibility is to rely on 
Monte Carlo simulations of heavy quark decays, for instance, on PYTHIA
\cite{ref:pythia}, which are quite successful and tuned to
a wealth of data. PYTHIA can be used to generate ``efficiencies'' $\varepsilon_{\mathrm{eff}}$
for observing a heavy quark within a certain bin in $p_T$ and $\eta$ in the detectors at RHIC.
If properly normalized to the total number of heavy quarks generated in that particular
bin, $\varepsilon_{\mathrm{eff}}$ should become independent of 
all the details of the heavy quark production mechanism assumed in PYTHIA.
This is essential for an unbiased extraction of $\Delta g$.

\begin{figure}[t]
\epsfxsize=\textwidth 
\epsfbox{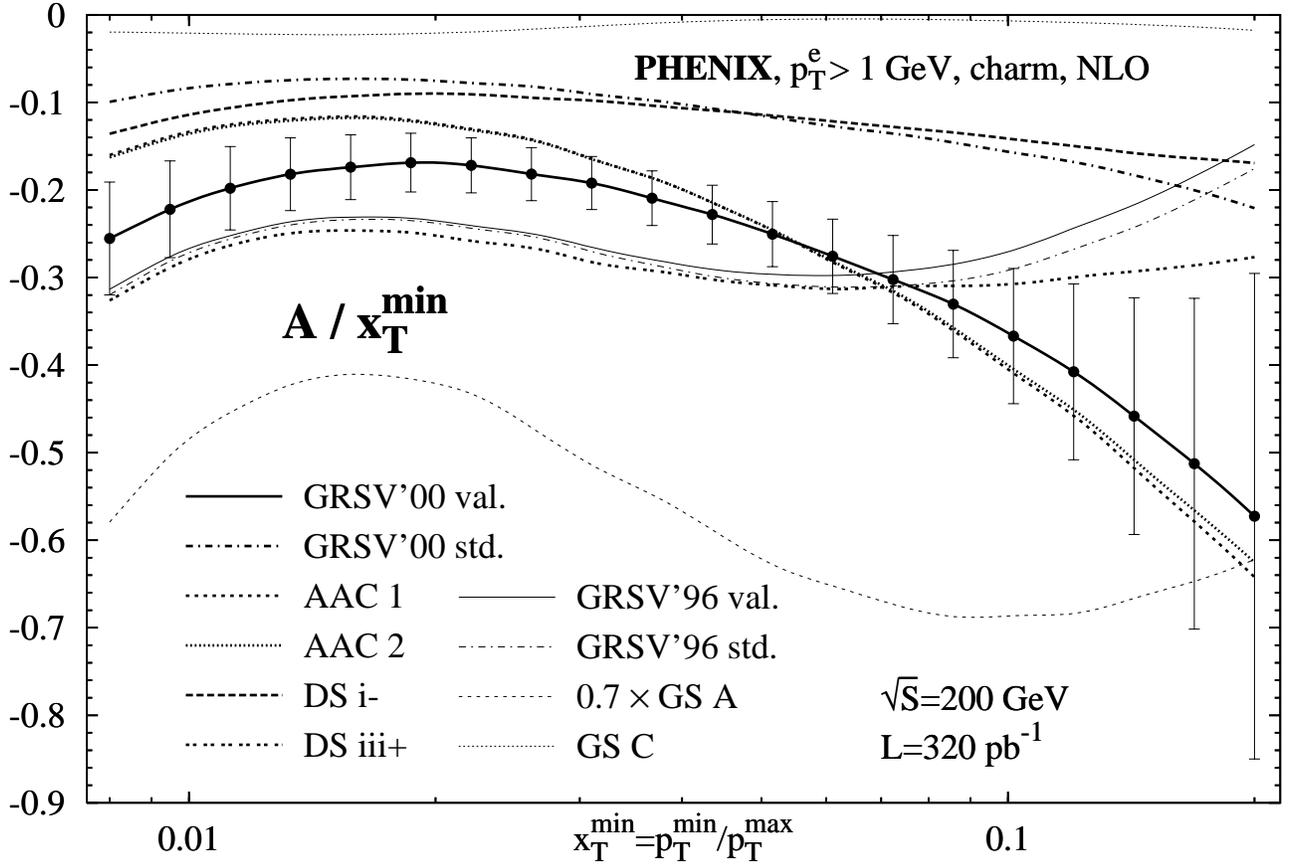}
\caption{\label{fig:casym} \sf
The NLO charm asymmetry $A$ at $\sqrt{S}=200\;\mathrm{GeV}$ for PHENIX at RHIC
as a function of $x_T^{\mathrm{min}}=p_T^{\mathrm{min}}/p_T^{\mathrm{max}}$
using Eq.~(\ref{eq:eef}).  
For a better separation of the curves $A$ is rescaled by $1/x_T^{\mathrm{min}}$.
Recent and old sets of helicity densities are distinguished by
thick and thin lines, respectively. 
Also shown is an estimate for the statistical error using 
a luminosity  of ${\cal L}=320\;\mathrm{pb}^{-1}$ (see text).}
\end{figure}

Exploiting this idea, a first numerical study for the PHENIX detector
has been performed\footnote{We are 
grateful to  M.\ Grosse Perdekamp for providing these efficiencies.}. 
The resulting efficiency $\varepsilon_{\mathrm{eff}}(p_T,\eta)$
for a charm quark produced with transverse momentum $p_T$ and 
pseudo-rapidity $\eta$ to be detected via its decay electron {\em anywhere} in the PHENIX 
acceptance, with the electron trigger allowing $p_T^e>1\,\mathrm{GeV}$,
is approximately given by
\begin{eqnarray}
\label{eq:peff}
&&\varepsilon_{\mathrm{eff}}(p_T,\eta;p_T^e>1\,\mathrm{GeV})=\zeta\;
     \exp\left(\frac{-9.79+4.58 (p_T/\mathrm{GeV})^{1.88}}
     {(p_T/\mathrm{GeV})^{1.73}+1.74\;\zeta^{-0.79}}\right)\;\;\mathrm{with}\\
&&\zeta=\exp\left\{-\left|\eta/\left(4.06\;
     \exp\left[-\left(p_T/1.05\;\mathrm{GeV}\right)^{0.43}\right]\right)
     \right|^{5.84\;\exp\left[-\left(p_T/2.48\;\mathrm{GeV}\right)^{0.42}\right]}
     \right\}\nonumber\;\;.
\end{eqnarray}
A prediction for the charm cross section as measurable with PHENIX is then
obtained by convoluting our double differential partonic results
with $\varepsilon_{\mathrm{eff}}$ in Eq.~(\ref{eq:peff}),
\begin{equation}
\tilde{\sigma}_{\mathrm{eff}}(p_T^e>1\,\mathrm{GeV})
=\int_0^{p_T^{\mathrm{max}}} dp_T
\int_{-\eta^{\mathrm{max}}}^{\eta^{\mathrm{max}}} d\eta\;
\varepsilon_{\mathrm{eff}}(p_T,\eta;p_T^e>1\,\mathrm{GeV})\;
\frac{d^2\tilde{\sigma}}{dp_Td\eta}\;\;,
\end{equation}
where $p_T^{\mathrm{max}}=\frac{1}{2}\sqrt{S-4 m^2}$ and
$\eta^{\mathrm{max}}=-\frac{1}{2}\ln\frac{1-\sqrt{1-4 p_T^2/(S-4 m^2)}}{1+\sqrt{1-4 p_T^2/(S-4 m^2)}}$ are the appropriate kinematical limits.
In the future \cite{ref:upcoming} we plan to use a set of
efficiencies with several different cuts on, or bins in, $p_T^e$ 
to generate different $\tilde{\sigma}_{\mathrm{eff}}$. For the time being,
different cuts in $p_T^e$ are simulated by limiting the charm transverse
momentum $p_T$ instead, while still using Eq.~(\ref{eq:peff}), i.e.,
\begin{equation}
\label{eq:eef}
\tilde{\sigma}_{\mathrm{eff}}(p_T^e>p_T^{\mathrm{min}})
\simeq\int_{p_T^{\mathrm{min}}}^{p_T^{\mathrm{max}}} dp_T
\int_{-\eta^{\mathrm{max}}}^{\eta^{\mathrm{max}}} d\eta\;
\varepsilon_{\mathrm{eff}}(p_T,\eta;p_T^e>1\,\mathrm{GeV})\;
\frac{d^2\tilde{\sigma}}{dp_Td\eta}\;\;.
\end{equation}
This expression has been used for the results shown in Fig.~\ref{fig:casym}.

To study the sensitivity of the charm production asymmetry at RHIC to $\Delta g$,
we use a range of recent \cite{ref:grsv,ref:aac,ref:ds} and old \cite{ref:grsv96,ref:gs}
helicity densities in Fig.~\ref{fig:casym}. These sets mainly differ in the
assumptions about $\Delta g$.
Note that for calculating the required unpolarized
$\sigma$ in $A=\Delta\sigma/\sigma$ we have used in each case
the underlying set of helicity averaged parton distributions as specified
in \cite{ref:grsv,ref:aac,ref:ds,ref:grsv96,ref:gs}. For consistency, $m$ is 
also taken as in these fits, i.e., $m=1.4\;\mathrm{GeV}$ for the modern 
and $m=1.5\;\mathrm{GeV}$ for the old spin-dependent densities.
All results are obtained for the choice $\mu_F^2=\mu_R^2=2.5 (m^2+p_T^2)$.
It is immediately apparent from Fig.~\ref{fig:casym} that charm production at
RHIC can be very useful in pinning down $\Delta g$. The estimated
statistical error for such a measurement,
$\delta A=\frac{1}{P_p^2}\frac{1}{\sqrt{{\cal L}\sigma_{\mathrm{eff}}}}$,
assuming a luminosity of ${\cal L}=320\;\mathrm{pb}^{-1}$ and a beam polarization 
of $P_p\simeq 0.7$ \cite{ref:rhic} is significantly smaller than the 
total spread of the predictions. This is true in particular if we 
take the large GS A \cite{ref:gs} $\Delta g$ into account, which is still compatible 
with recent data. The GS A asymmetry is so large that it
had to be scaled down by 0.7 to fit well into the same plot. 
Note that very small gluons, e.g., the oscillating $\Delta g$ of
GS C \cite{ref:gs}, are at the edge of being detectable.
There are three groups of helicity densities which will be indistinguishable
within the errors by charm production: GRSV'00 std.\ \& DS i-, AAC 1 \& 2, and 
DS iii+ \& GRSV'96 std. \& val. Here the gluon densities are too similar in the range of
$x$ predominantly probed by this process. 
We will map the range in $x$ where $\Delta g$ is accessible by heavy flavor production
at RHIC in detail in \cite{ref:upcoming}. 

\begin{figure}[th]
\epsfxsize=\textwidth 
\epsfbox{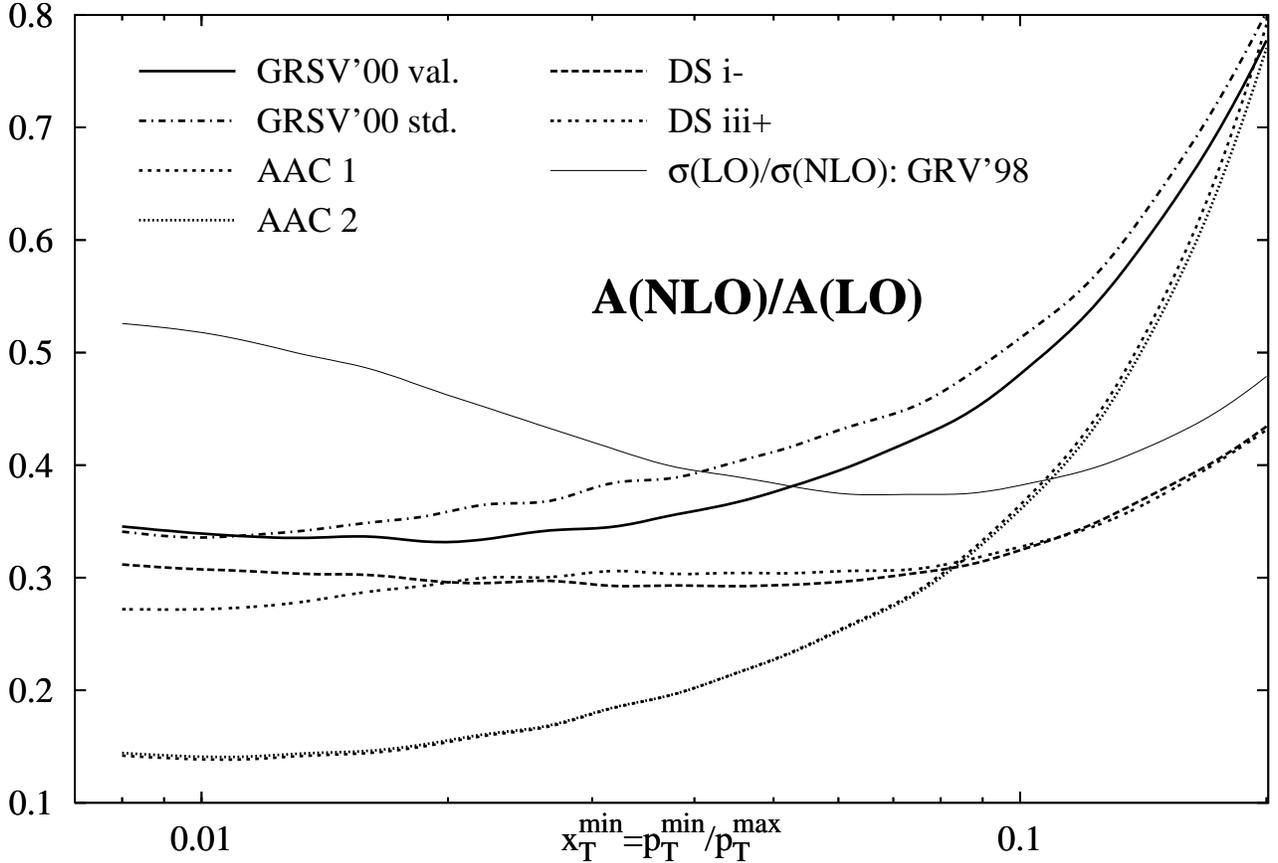}
\caption{\label{fig:kfac}\sf Ratio of the asymmetries in NLO and LO,
$A(\mathrm{NLO})/A(\mathrm{LO})$, with 
$A(\mathrm{NLO})$ as shown in Fig.~\ref{fig:casym}. 
The ratio of unpolarized cross sections
$\sigma(\mathrm{LO})/\sigma(\mathrm{NLO})$ used in the calculation of $A$
is also shown for comparison (thin solid line).}
\end{figure}
Finally we take a look at the importance of the NLO corrections for the
heavy flavor spin asymmetry. Fig.~\ref{fig:kfac} shows the ratio 
of the charm spin asymmetries calculated in NLO (as shown in Fig.~\ref{fig:casym})
and LO for the range of modern helicity densities \cite{ref:grsv,ref:aac,ref:ds}.
One can infer that the NLO asymmetries are generally {\em smaller} than the 
LO ones by a factor of about three. 
In the case of the AAC helicity densities \cite{ref:aac} we find an even larger suppression. 
Furthermore, there is considerable dependence on $x_T^{\mathrm{min}}$,
which will inhibit the use of constant ``K-factors'' to estimate the NLO results
from LO ones.
It should also be pointed out that much of the reduction of the asymmetry
in NLO stems from the {\em unpolarized} cross section in NLO, which is about a factor
two larger than the corresponding LO result.
This is shown by the thin solid line in Fig.~\ref{fig:kfac} representing
the ratio $\sigma(\mathrm{LO})/\sigma(\mathrm{NLO})$ obtained with
the GRV'98 parton densities \cite{ref:grv98}. 
The sizable difference of the asymmetry predictions in LO and
NLO means that the LO and NLO gluon helicity densities extracted from
a future asymmetry measurement will differ considerably. 
Whether this will be consistent with data from
other processes has to studied in a global analysis, e.g.,
along the lines suggested in Ref.~\cite{ref:mellin}.

Further studies of the uncertainties and predictions for bottom production will
become available soon \cite{ref:upcoming}. There we will also present more details
concerning the calculational techniques that have been used and analytical results
for the matrix elements that we have obtained. 
In case one wants to extend our gluon-gluon results 
to the production of gluinos, one has to take into account the following colorfactor
replacements in Eqs.~(\ref{eq:mvgg}) and (\ref{eq:mrgg}), 
which already include the color-average:
replace the prefactor $\frac{1}{2(N_C^2-1)}\to\frac{N_C}{N_C^2-1}$ and set
$C_F=C_A=N_C$ inside the square brackets. Before doing the latter, one has to
use the identity $1=C_A^2-2 C_FC_A$ for the {\em KQ} parts, e.g.,
$\tilde{U}_{KQ}=(C_A^2-2 C_FC_A)\tilde{U}_{KQ}\to -N_C^2\tilde{U}_{KQ}$.
We note that one could use this identity anyway
to rewrite the results with the same number of colorfactors, e.g., 
$C_A^2\tilde{U}_{OQ}+\tilde{U}_{KQ}=C_A^2(\tilde{U}_{OQ}+\tilde{U}_{KQ})-2 C_F C_A\tilde{U}_{KQ}$,
but this has numerical disadvantages due to the length of the expressions for
$\tilde{U}_{OQ}$ and $\tilde{U}_{KQ}$. For a complete NLO supersymmetric QCD calculation of
spin-dependent gluino production one would
have to take into account that the running of the strong coupling
will be changed and that other subprocesses will contribute. 
However, as always, gluon-gluon scattering is expected to be the dominant subprocess. 

%
\section{Summary}
\noindent
To summarize, we have presented the first complete NLO QCD calculation for the
spin-dependent hadroproduction of heavy quarks. The NLO
results have considerably less uncertainties stemming from variations of
the unphysical factorization and renormalization scales. They even become nearly
independent of the scales when the conventional choice $\mu_F=\mu_R$ is
employed. There remains, however, an uncertainty of $\Delta\sigma$ of about $\pm30\%$ 
from variations of the charm mass within a reasonable range. 
We have made predictions for the charm asymmetry that can be measured soon with 
the PHENIX experiment at RHIC. These predictions include an efficiency
function which describes hadronization, decays of the heavy quarks, 
experimental cuts, and the detector geometry. 
In this way it was demonstrated clearly that $\Delta g$ can be constrained
considerably with heavy flavor production, even if current experimental 
limitations are taken into account realistically. 
Finally, we have shown that as in the unpolarized case, LO calculations 
cannot be substituted in any simple manner for the full NLO result. 
Our work is of particular relevance also to the ``heavy quark enigma'' 
arising from discrepancies of unpolarized bottom, but not charm, production 
data with theory. On the one hand
one now will be able to compare {\em spin-dependent}
RHIC data for charm and bottom production to a NLO prediction. 
On the other hand our calculation provides the major part of 
the spin-dependence of a possible ``new physics'' explanation 
of this discrepancy in terms of supersymmetry by a simple adjustment of colorfactors.
More details and further phenomenological studies can be found in \cite{ref:upcoming}.

%
\section*{Acknowledgements}
\noindent
We are grateful to M.\ Grosse Perdekamp  and J.\ Smith for helpful discussions.
M.S.\ thanks SUNY Stony Brook, RIKEN, and Brookhaven National
Laboratory for hospitality and support during the final steps
of this work. I.B.'s work has been supported by
the ``Bundesministerium f\"{u}r Bildung, Wissenschaft,
Forschung und Technologie'' and the Australian Research Council.
%

%
\end{document}